\documentclass[twocolumn,showpacs,preprintnumbers,amssymb]{revtex4}
\usepackage{amsmath2000}
\usepackage{epsfig}
\usepackage{bm}
\usepackage{dcolumn}

\begin{document}
\preprint{Phys.\ Rev. E {\bf 65} (2002)}
\title{Electric microfield distributions in electron-ion plasmas}

\author{Alexander Y. Potekhin}\email{palex@astro.ioffe.rssi.ru}
\affiliation{Ioffe Physical-Technical Institute,
     194021 St.\ Petersburg, Russia}
\author{Gilles Chabrier}\email{chabrier@ens-lyon.fr}
\affiliation{Ecole Normale Sup\'erieure de Lyon,
     CRAL (UMR CNRS No.\ 5574),
     69364 Lyon Cedex 07, France}
\author{Dominique Gilles}\email{Dominique.Gilles@cea.fr}
\affiliation{Commissariat \`a l'Energie Atomique,
BP 12, 91680 Bruy\`eres-le-Ch\^atel,
 France}

\date{5 July 2001, accepted 17 December 2001}

\begin{abstract}
The low-frequency electric microfield distribution in a Coulomb
plasma is calculated for various plasma parameters,
from weak to strong Coulomb coupling
and from zero to strong electron screening.
Two methods of numerical calculations are employed:
the adjustable-parameter exponential approximation
and the Monte Carlo simulation.
The results are represented by analytic fitting formulas
suitable for applications.
\end{abstract}



\maketitle

\section{Introduction}
Because of the Stark effect,
stochastic electric microfields influence optical and
thermodynamic properties of a plasma.
First, they affect the profiles of spectral lines
and effectively lower photoionization thresholds
of atoms and ions immersed in a plasma \cite{Griem,Fuhr}.
A comparison of experimental and theoretical widths and shapes
of the Stark-broadened spectral lines is widely used
for plasma diagnostics
(e.g., Refs.~\cite{Jiang,Vitel}).
Second, in some theoretical models of the plasma equation
of state (e.g., Refs.~\cite{HM,Nayfonov}),
the microfield distribution is used in order to
calculate occupation numbers of
the bound species (although such a calculation is not free
from principal difficulties, as discussed in Ref.~\cite{P96}).
It was shown recently \cite{Nayfonov} that
a more accurate description of the microfields entails
a considerable improvement of the equation-of-state model.

In many cases,
the microfield perturbation can be treated as quasistationary.
Then the problem is reduced to determination of the probability
distribution of the \textit{low-frequency} component
of perturbing electric fields (e.g., Ref.~\cite{APEX2}),
associated with a stochastic distribution of perturbing ions,
whereas the electrons can be assumed to adjust
instantaneously to a configuration of the ions.
The low-frequency microfields are appropriate to use
in the equation of state models \cite{Nayfonov} and
in calculation of spectroscopic line profiles for those
radiative transitions whose frequency does not exceed
the typical frequency of microfields produced by
thermal fluctuations of the electron density.
For example, Stehl\'e and Jacquemot \cite{SJ93}
used the model microfield method to
analyze the line shapes and line
dissolution in hydrogen plasma spectra.

Holtsmark \cite{Holtsmark} has derived the microfield distribution
function assuming that the ions are not correlated and the
electron screening is negligible.
This assumption is justified for very
hot or rarefied plasmas, for which the Coulomb coupling parameter
\begin{equation}
   \Gamma = \frac{(Ze)^2}{a k_B T}
      \approx\frac{1.25\times10^4{\rm~K}}{T}\,n_{20}^{1/3}\,Z^{5/3}
\end{equation}
is close to zero. Here,
$Ze$ is the ion charge, $T$ is the temperature,
$k_B$ is the Boltzmann constant,
$a=(\frac{4\pi}{3}n_i)^{-1/3}$
is the ion sphere radius, $n_i$ is the ion number density, and
$n_{20}$ is the electron number density ($n_e=Zn_i$)
in units of $10^{20}{\rm~cm}^{-3}$.
As we demonstrate below, the Holtsmark approximation
is inaccurate already at $\Gamma\sim0.1$.
In modern plasma experiments, $\Gamma$ may approach unity,
whereas in stellar matter it can be much larger.
In these cases,
correlations of plasma particles should not be neglected.

Various approximations were developed in the past in order
to take the ion correlations into account.
If $\Gamma\lesssim1$, one may use the methods
of Baranger and Mozer \cite{BM} or Hooper \cite{Hooper,Hooperasymp}
based on a cluster expansion in powers of density.
The electron screening is usually described by a Debye-like (Yukawa)
effective potential,
introduced in the context of microfield distributions
by Hoffman and Theimer \cite{HoffmanTheimer}.
In the limit of extremely strong coupling, $\Gamma\gg10$,
and without screening,
the harmonic oscillator model by Mayer \cite{Mayer} is applicable,
in which every ion is assumed to oscillate independently of the others
around its equilibrium position at the ion-sphere center.

The first theory capable to provide reliable numerical
results for strongly coupled plasmas with electron screening
proved to be the adjustable-parameter exponential approximation (APEX),
based on a special parametrization
of the electric microfield $\bm{E}$
produced on a selected test particle (neutral or charged ``radiator'' of
charge $Z_r$)
 which undergoes the influence of charged plasma particles
(``perturbers'' of species $\sigma$ and of charge $Z_{\sigma}$).
This method has been developed for Coulomb systems \cite{APEX} and
adapted for screened Coulomb systems and ion mixtures
\cite{APEX-multi,APEX2}.
It involves non-interacting quasiparticle representation of the
electron-screened ions, designed to yield the correct second moment
of the microfield distribution \cite{Iglesias&al}:
\begin{equation}
\langle \bm{E}\cdot\bm{E} \rangle=
{4\pi n_i k_B T\over Z_r} k_s^2 \sum_\sigma c_\sigma Z_\sigma
\int_0^\infty dr\,r e^{-k_s r} g_\sigma(r),
\label{EE}
\end{equation}
where $g_\sigma(r)$ and  $c_\sigma$ denote the radial distribution
function (RDF) and the relative abundance of species $\sigma$,
respectively, and where $k_s$ is an effective electron screening
wave-number.
After introducing the effective single-particle field in the form
\begin{equation}
\epsilon^{*}_{\sigma}=Z_{\sigma} e {(1+ \alpha_\sigma r) \over r^2}
\,e^{-\alpha_{\sigma}r},
\end{equation}
the adjustable parameters $\{\alpha_\sigma \}$ are chosen to satisfy
the condition
\begin{equation}
4\pi \int \epsilon^2 P_\textrm{APEX}(\epsilon) d\epsilon =
\langle \bm{E}\cdot\bm{E} \rangle.
\end{equation}
The expression on the left-hand side of this equation
contains the parameters $\{ \alpha_{\sigma } \}$ to be determined,
whereas the right-hand side can be evaluated using Eq.~(\ref{EE}),
if the RDF is known.
The RDF thus provides a scheme for evaluating the APEX microfield
distribution and is a central ingredient whose accuracy determines
the one of the APEX microfield results.
In our implementation of the APEX technique,
we have used the hypernetted-chain
RDF calculations \cite{Iglesias&al,Rogers,Chabrier}.

On the other hand, with the advent of powerful computers
it is now possible to calculate the microfield
distribution from Monte Carlo (MC) or molecular-dynamics
simulations of plasmas
with the minimum of simplifying assumptions (e.g.,
Refs.~\cite{AG86,Stamm,GS95,Gilles97,CG00}).
Moreover, the latter methods allow one to study the effects of
microfield nonuniformity \cite{Demura,Murillo}
and to simulate high-frequency microfield distributions
in electron-ion plasmas (e.g., Ref.~\cite{Filinov}).
The MC technique is based on a numerical simulation of
space configurations of a system of particles,
whereas the molecular-dynamics technique
traces the time evolution of the system.
For the low-frequency microfield,
dynamical effects are unimportant,
and the two methods yield the same results, as
demonstrated, e.g., in Ref.~\cite{Murillo}.
Therefore it is sufficient to use the MC method in this case.

With these powerful tools,
the microfield distribution can be calculated
now for any practically important combination of plasma parameters.
However, plasma spectroscopy and equation-of-state models
require knowledge of this distribution at many different
points or even in continuous areas of the plasma parameter space.
In this case, either extensive numerical tables
or approximate analytic expressions are necessary.

We present results of calculations of the
low-frequency microfield distribution
function at a neutral and charged plasma point
for various values of $\Gamma$ ranging from 0 to 100
 and for various values of
an effective electron screening length.
We consider plasmas composed of a single species of ions;
in particular, in the case of a charged test particle, its charge
is assumed to be equal to that of perturbers.
The calculations are performed mainly by the MC method;
for comparison we have done also APEX calculations.
We also present analytic formulas which reproduce
the calculated electric microfield probability distributions
with an accuracy comparable to small differences between the MC
and APEX results.

In the next section, we describe basic assumptions used
in our calculations and write down some asymptotic results.
In Sec.~\ref{sect-res}, we present results of numerical calculations and
analytic approximations for microfield distributions
produced at a neutral or charged point by ions interacting
via unscreened or screened Coulomb potentials.
The results are summarized in Sec.~\ref{sect-end}.

\section{Method}

\subsection{Basic assumptions}

We consider a nonrelativistic,
isotropic, overall neutral plasma at the thermodynamic equilibrium.
The ions are assumed to be classical and pointlike.
The electric field created at a point $\bm{r}$
by an ion placed at $\bm{r}_i$ equals
$\tilde{\bm{E}}(\bm{r}_i-\bm{r})=-(Ze)^{-1}\nabla V(|\bm{r}_i-\bm{r}|)$,
where $V(r)$ is an effective pair potential.
This potential is taken in the Debye--H\"uckel (or Yukawa) form:
\begin{equation}
   V(r) = (Ze)^2\,{e^{-k_s r}\over r}.
\label{V}
\end{equation}
In the linear approximation,
a test charge $q$ embedded (at $\bm{r}=0$)
creates perturbation of electron number
density $\tilde n_e(r)=(\partial n_e/\partial\mu) \tilde\mu(r)$, where
$\tilde\mu(r)=-e\phi(r)$ is the perturbation of the electron
chemical potential $\mu$, and
$\phi(r)$ is the excess electrostatic potential
determined by the Poisson equation
$\nabla^2 \phi(r) = -4\pi [q\delta^3(\bm{r})-e\tilde n(r)]$.
Thus in the linear (first-order perturbation) approximation
$(\nabla^2 + 4\pi e^2\partial n_e/\partial\mu)\phi(r)
 = -4\pi q \delta^3(\bm{r})$, which leads to the well-known
(e.g., Ref.~\cite{Chabrier}) expression
for the effective screening wave number $k_s$:
\begin{equation}
   k_s^2 = 4\pi e^2\,{\partial n_e\over\partial\mu}
    =  {e^2\over\pi\hbar^3}\,(2m_e)^{3/2} \,(k_B T)^{1/2}
       \,I_{-1/2}(\chi),
\label{k_TF}
\end{equation}
where
$I_\nu(\chi)=\int_0^\infty x^\nu\,dx/(e^{x-\chi}+1)$
is the Fermi integral,
and $\chi\equiv\mu/k_B T$ is determined from the equation
\begin{equation}
    I_{1/2}(\chi)=\pi^2\hbar^3\,(m_e\,k_B\,T)^{-3/2}\,n_e/\sqrt{2}.
\label{chi}
\end{equation}
The solution of Eq.~(\ref{chi}) and the right-hand side of Eq.~(\ref{k_TF})
are given by accurate Pad\'e approximations in Ref.~\cite{Antia}.

In the limits of weak or strong electron degeneracy, $k_s$
tends to the inverse Debye length for the electrons
or to the Thomas-Fermi wave number, respectively.

We adopt the conventional assumption that the potentials
$V(|\bm{r}_i-\bm{r}|)$ are additive, which is strictly
valid in the limit $k_s\to0$. Then the electric field $\bm{E}(\bm{r})$ is also
the sum of elementary electric fields $\tilde{\bm{E}}(\bm{r}_i-\bm{r})$.

It is convenient to introduce the dimensionless field $\bm{\beta}$
and the screening parameter $s$:
\begin{equation}
   \bm{\beta}=(a^2/Ze)\,\bm{E},
\qquad
   s=a\,k_s.
\label{param}
\end{equation}
Also, $\tilde{\bm{\beta}}=(a^2/Ze)\,\tilde{\bm{E}}$.

In the canonical thermodynamic ensemble of $(N+1)$ particles,
the probability density
of the modulus of the field, $\beta=|\bm{\beta}(\bm{r}_0)|$,
can be written as
\begin{eqnarray}
  P(\beta) & = & {4\pi\beta^2\over {\cal Z}_{N+1}}
   \int\ldots\int
   \delta\left[{\bm{\beta}}(\bm{r}_0)
       -\sum_{i=1}^N\,\tilde{\bm{\beta}}(\bm{r}_i-\bm{r}_0)\right]
\nonumber\\&&\times
        e^{-W(\bm{r}_0,\bm{r}_1,\ldots,\bm{r}_N)/k_B T}
 d\bm{r}_1\ldots d\bm{r}_N.
\label{P-gen}
\end{eqnarray}
Here,
${\cal Z}_{N+1}=\int\ldots\int
e^{-W(\bm{r}_0,\bm{r}_1,\ldots,\bm{r}_N)/k_B T}
 d\bm{r}_1\ldots d\bm{r}_N$ is the canonical partition function
and $W$ is
the potential energy of the configuration:
$W(\bm{r}_0,\bm{r}_1,\ldots,\bm{r}_N)=
       \frac12 \sum_{i\neq j} V(|\bm{r}_i-\bm{r}_j|) + V_B$,
 where $V_B$ is the potential energy of the background of electrons.
Our goal is to calculate the function $P(\beta)$
in the thermodynamic limit $N\to\infty$. In the next paragraph we shall
discuss how to perform this calculation.

\subsection{Monte Carlo technique}

In the numerical MC calculations,
the coordinates of $(N+1)$ particles
(one test particle and $N$ perturbers)
are chosen in a cubic box of side length $L$
such that $(N+1)/L^3=n_i$.
In order to include the effect of distant particles,
the box is replicated by its ``images''
filling the space with the step $L$.
The sum of the interaction
potentials with all ion images is calculated using
the Ewald technique \cite{Ewald}
and, in order to ensure isotropy,
the averaging over the complete solid angle is applied.
As shown in Ref.~\cite{AG86}, a result of this procedure
is equivalent to the replacement of the potential (\ref{V})
in the cube by an effective potential $V_L$,
\begin{equation}
   {V_L(r)\over(Ze)^2}= {e^{-k_s r}\over r}
       - {C_M'\over L}\,{\sinh(k_s r)\over k_s r}
       + {4\pi\over k_s^2 L^3}\left[{\sinh(k_s r)\over k_s r}-1\right],
\end{equation}
where $C_M'$ is a numerical constant
which tends to the Madelung constant (e.g., Ref.~\cite{brush})
in the limit $k_s\to0$.
The total potential energy $W$ is obtained by the summation of
$V_L$ over all pairs of particles.

During the MC run, an ion within the box
and its displacement are chosen
randomly. If the displacement brings the ion
outside the box, the ion is replaced by its image.
In strongly correlated plasmas, the displacement
is limited by a maximum distance smaller than $L$,
in order to avoid calculation of the energy
for highly improbable configurations.
First $10^4$ ion configurations are discarded
in order to erase traces of the starting configuration.
The state of equilibrium is searched by the Metropolis algorithm:
the energy difference $\Delta W$ is calculated
between the consecutive configurations, and the new configuration
is accepted definitely if this difference is negative
and accepted with probability $e^{-\Delta W/k_B T}$
if $\Delta W$ is positive. The latter condition allows the system
to escape from trapping in a local energy minimum.
When the system approaches
equilibrium, $W$ ceases to change appreciably.
Then all equilibrium quantities depending only
on particle positions (electric microfield,
pair correlation function, etc.) can be calculated.
In all our simulations, we try to get the maximum precision by
minimizing the statistical errors. So we considered large boxes of
particles (between $N=600$ and $N=800$)
and, for each state $(\Gamma, s)$, (1--$6)\times10^7$ configurations
were generated after equilibrium. This number of samples is large enough
for the precision needed for all microfield distribution results discussed
in this paper.

\subsection{Asymptotic and approximate expressions}
\label{sect-asymp}

The described MC sampling procedure does not
directly provide the probability density for
extremely weak or strong fields, which are given by
rare configurations. It is therefore useful
to know asymptotic behavior of $P(\beta)$
in the limits $\beta\to0$ and $\beta\to \infty$.

In the first case, the methods of Baranger and Mozer \cite{BM}
and APEX \cite{APEX2,APEX} show a parabolic dependence
of $P(\beta)$ near the origin. This behavior is also visible on MC results.
Thus we assume that $P(\beta)/\beta^2$ is constant near the origin.

In the case of very strong fields,
exact analytic results are available
for the unscreened Coulomb potential only.
The Holtsmark distribution, valid at $\Gamma\to0$ for any $\beta$,
reads \cite{Holtsmark}
\begin{equation}
   P_H(\beta)={2\beta\over\pi}\int_0^\infty x\,
   \exp(-x^{3/2})\sin(\beta x) \,dx.
\label{Holts}
\end{equation}
At $\beta\to\infty$, this distribution has the asymptote
$P(\beta)\sim 1.496\,\beta^{-5/2}$,
which is close to the asymptote of the
nearest-neighbor (NN) field distribution
\cite{LM58}
\begin{equation}
   P_\textrm{NN}(\beta)=1.5\,\beta^{-5/2}\,\exp(-\beta^{-3/2}).
\label{NN}
\end{equation}

In the opposite limit of extremely strong correlations
($\Gamma\to\infty$), the Mayer model \cite{Mayer} yields
(for the charged test particle)
\begin{equation}
   P_M(\beta)= \sqrt{2/\pi}\;\Gamma^{3/2}\,\beta^2\,
      \exp(-\Gamma\beta^2/2).
\label{Mayer}
\end{equation}

\begin{figure}
    \begin{center}
    \leavevmode
    \epsfxsize=86mm
    \epsfbox[145 280 460 590]{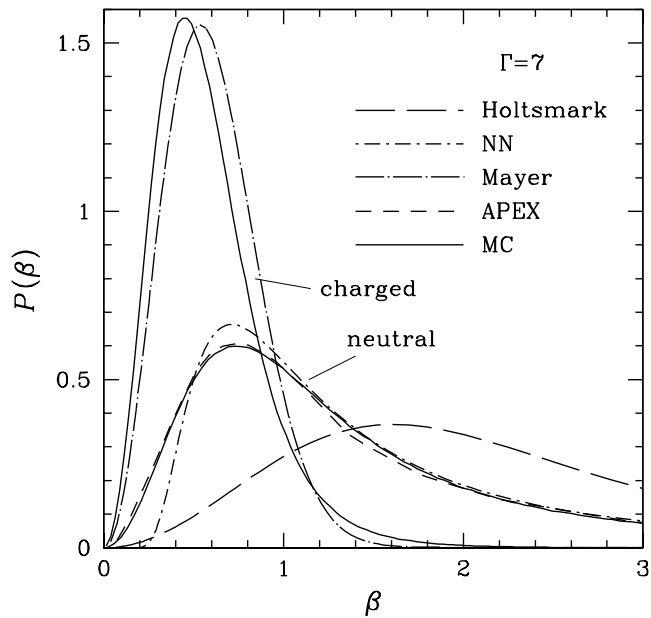}
    \end{center}
\caption{
Microfield distributions produced by ions
interacting via the unscreened Coulomb potential
in a plasma at $\Gamma=7$
at a neutral and charged point in the plasma.
MC (solid lines)
and APEX (short-dashed curve) numerical results are compared with
analytic approximations by Holtsmark (long dashes),
Mayer and nearest neighbor (dot-dashed lines).
}
\label{fig-comp0}
\end{figure}

Figure \ref{fig-comp0} illustrates
the differences between
various asymptotic theories (NN, Mayer model) and numerical results
at finite $\Gamma$.
Compared to the Holtsmark distribution (\ref{Holts}),
the most probable field
values are shifted considerably to lower $\beta$,
the shift being much larger at a charged point.
The NN approximation (\ref{NN}) correctly describes
the case of large $\beta$ for the neutral point but fails
for the charged point or at small $\beta$.
The Mayer distribution (\ref{Mayer})
fails to describe the high-field tail of $P(\beta)$ but
provides
the most probable field which, at this $\Gamma$,
is offset by tens percent only.
At contrast, the APEX and MC results are in close agreement
for $\beta <2$.

\begin{figure}
    \begin{center}
    \leavevmode
    \epsfxsize=86mm
    \epsfbox[130 240 460 560]{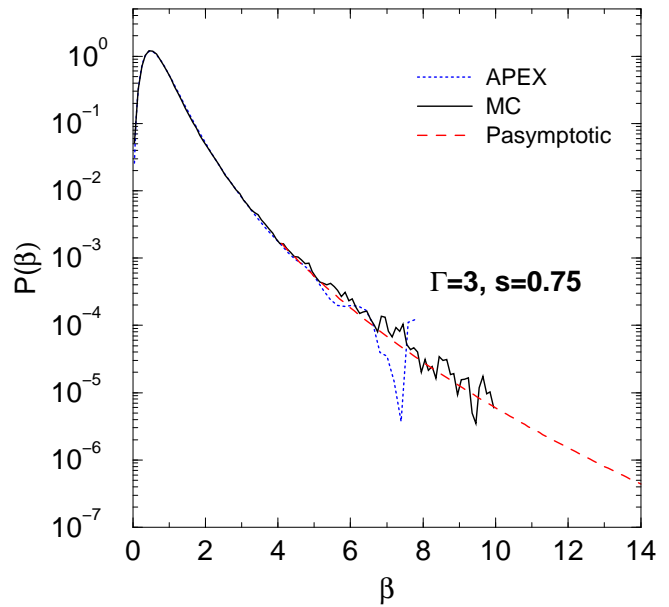}
    \end{center}
\caption{
Comparison, for $\Gamma=3$ and $s=0.75$,
of probability density distributions $P(\beta)$ at a charged point,
calculated by different methods:
Monte Carlo (solid line), APEX (dotted line),
and the asymptotic expression (\ref{Pasymp})
(dashed line). Coefficients of the asymptotic expression have been
fitted on MC points above the cut-off value $\beta=5$:
 $\tilde\Gamma=3.06$ and $\tilde K=31.011$.
 The asymptotic expression is a good representation of the
high-field tail of $P(\beta)$
 since it avoids the oscillations
shown by the two other methods (MC: numerical statistical noise,
 APEX: Fourier transform oscillations).
}
\label{fig-microMCapexASYMP}
\end{figure}
\begin{figure}
    \begin{center}
    \leavevmode
    \epsfxsize=86mm
    \epsfbox[145 280 460 590]{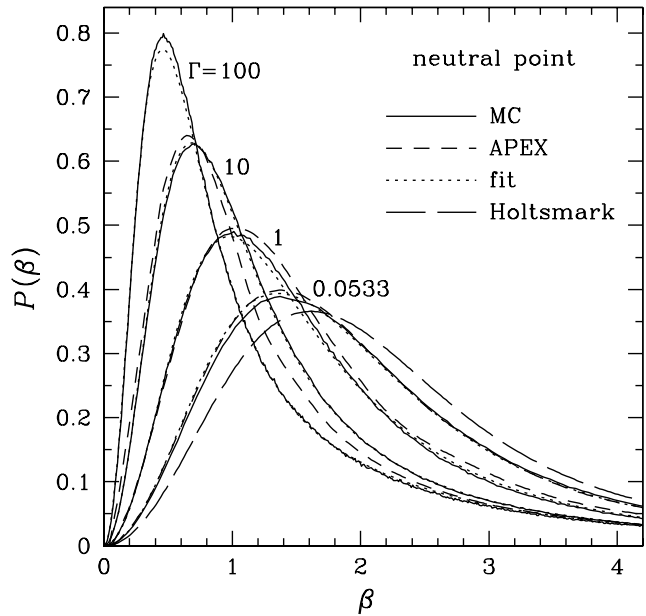}
    \end{center}
\caption{
Microfield distributions produced
at a neutral point by ions
interacting via the Coulomb potential,
for $\Gamma=0.0533$, 1, 10, and 100. Numerical results
(MC, solid lines; APEX, short-dashed lines) are compared with
analytic approximation (\protect\ref{fit-micron}) (dotted curves).
The long-dashed curve reproduces the
Holtsmark distribution (\protect\ref{Holts}).
}
\label{fig-micron0}
\end{figure}

The Mayer model fails in the strong-field limit, because
in this case one should consider
a test ion which lies at a very short distance $r\propto\beta^{-1/2}$
from the nearest perturbing ion.
Then the geometrical and Boltzmann factors
give $P(\beta)\sim \beta^{-5/2}\,e^{-\Gamma\beta^{1/2}}$
at $\beta\to\infty$. In Appendix~\ref{sect-C} we present derivation of
a more accurate asymptotic expression, Eq.~(\ref{Paris}), which
was previously given in Refs.~\cite{Gilles97,Paris} for the case of
the Coulomb potential without screening.
A generalization of Eq.~(\ref{Paris})
provides an accurate functional form
of the asymptotic behavior of $P(\beta)$ at large $\beta$
\cite{Gilles97,Gh95}:
\begin{equation}
   P(\beta) \sim
    \tilde K\,\beta^{-5/2}\,\exp(-\tilde\Gamma\beta^{1/2}-\beta^{-3/2}),
\label{Pasymp}
\end{equation}
where $\tilde K$ and $\tilde\Gamma$ are adjustable parameters.
In practice, $\tilde\Gamma$ is a free parameter,
whereas $\tilde K$ can be determined from the normalization constraint.
An example
is given in Fig.~\ref{fig-microMCapexASYMP}.
In this example, the value of
$\tilde\Gamma$ is close to the exact $\Gamma$, but this is not the general
case. However, as shown in Refs.~\cite{Gilles97,Gh95}, for any simulation
of practical interest, it is always
possible to find  appropriate $\tilde\Gamma$ and $\tilde K$
to fit the high-field tail of the microfield distribution.

\section{Results}
\label{sect-res}

\subsection{Coulomb potential}
\label{sect-Coulomb}

In this section, we present results of MC and APEX
calculations and analytic approximations for
the microfield distribution
brought about by the one-component plasma ions
interacting via the Coulomb potential without screening.
In this case, the probability density
 $P$ depends on $\beta$ and $\Gamma$.

\subsubsection{Neutral point}

First we consider the distribution of electric microfields
applied to a neutral test particle embedded in a plasma.
If the plasma is weakly coupled, $P(\beta)$ is given by the Holtsmark
formula (\ref{Holts}).
In theoretical models (e.g., \cite{HM,Nayfonov})
one is often interested in the cumulative probability distribution
defined as
\begin{equation}
   Q(\beta) = \int_0^{\beta} P(\beta')\,d\beta'.
\label{Q}
\end{equation}
For the Holtsmark distribution,
 $Q_H(\beta) \approx 1-0.997\beta^{-3/2}$
at $\beta\to\infty$.
For arbitrary $\beta$, accurate rational-function approximations
to $P_H(\beta)$ and $Q_H(\beta)$
have been constructed by Hummer \cite{Hummer}.

With increasing $\Gamma$, the field distribution becomes narrower,
as shown in Fig.~\ref{fig-micron0}.
The decrease of the most probable value $\beta_m$
of the dimensionless field $\beta$,
which corresponds to the maximum of $P(\beta)$,
can be described by a simple approximate formula
\begin{equation}
   \beta_m^\textrm{neu}\approx
     {1.608 + 0.24\,\sqrt{\Gamma} \over 1+0.77\,\sqrt{\Gamma}}.
\label{beta_m-neu}
\end{equation}
The asymptotic behavior of $P(\beta)$ remains power-law,
as in the case without Coulomb coupling.
This facilitates construction of self-consistent rational
approximations to $Q(\beta)$ and $P(\beta)$.
We have calculated $P(\beta)$ for various $\Gamma$ from 0 to 10
by the APEX method and for $\Gamma$ up to 100 by the MC technique.
Our fit to $Q(\beta)$ reads
\begin{equation}
   Q(\beta)={q_0\,\beta^3-1.33\,\beta^{9/2}+\beta^6
       \over q_1+q_2\,\beta^2+q_3\,\beta^3-\frac13\beta^{9/2}+\beta^6},
\label{fit-micron}
\end{equation}
where
$
   q_n = \alpha_n\,(1+\beta_n\,\sqrt{\Gamma})^{-\gamma_n},
$
and the parameters $\alpha_n$, $\beta_n$, and $\gamma_n$
are given in Table~\ref{tab-micron}. $P(\beta)$
is obtained from Eq.~(\ref{fit-micron}) by elementary differentiation.
At $\Gamma=0$, this differentiation
reproduces $P_H(\beta)$ at any $\beta$
with a maximum fractional error of 0.24\%.
At finite $\Gamma$, the difference between the fit
and the MC data
increases up to several percent, remaining however
not larger than the difference
between the MC and APEX results,
as one can see in Fig.~\ref{fig-micron0}.

\begin{table}
\caption{Parameters of Eq.~(\protect\ref{fit-micron}).}
\label{tab-micron}
\begin{ruledtabular}
\begin{tabular}{cdddd}
  $n$ & 0 & 1 & 2 & 3 \\
\noalign{\smallskip}
\hline
\noalign{\smallskip}
$\alpha_n$ & 14.600 & 103.20 & 11.127 & 16.178 \\
$\beta_n$  & 0.41   & 1.54   & 0.58   & 0.60 \\
$\gamma_n$ & 0.707  & 1.64   & 0.572  & 0.915 \\
\end{tabular}
\end{ruledtabular}
\end{table}

\begin{figure}
     \begin{center}
     \leavevmode
     \epsfxsize=86mm
     \epsfbox[145 280 460 640]{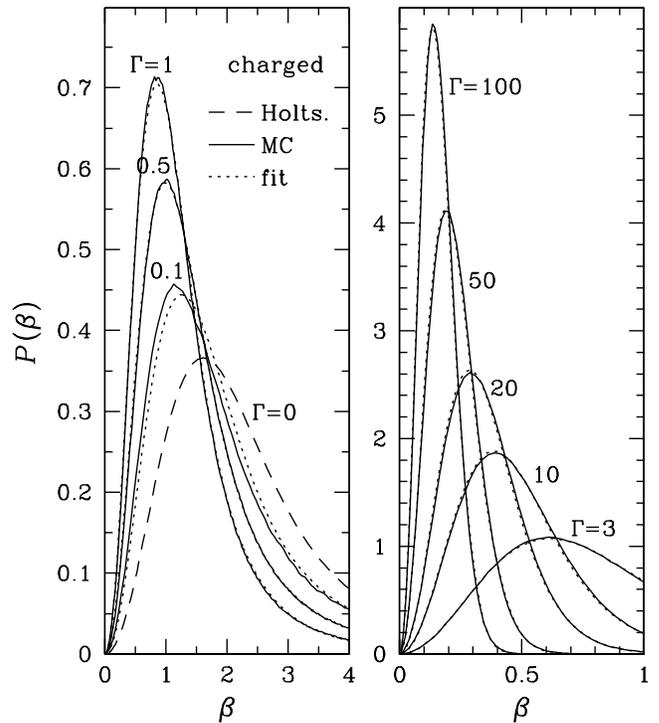}
     \end{center}
\caption{
Microfield distributions produced
at a charged test particle by ions
interacting via the Coulomb potential
at various $\Gamma$ from 0.1 to 100.
MC results (solid lines)
are compared with
analytic approximation (\protect\ref{fit-microf}) (dotted curves).
The Holtsmark distribution ($\Gamma=0$)
is also plotted (dashed line).
}
\label{fig-microf0}
\end{figure}
\begin{figure}
     \begin{center}
     \leavevmode
     \epsfxsize=86mm
     \epsfbox[150 280 460 500]{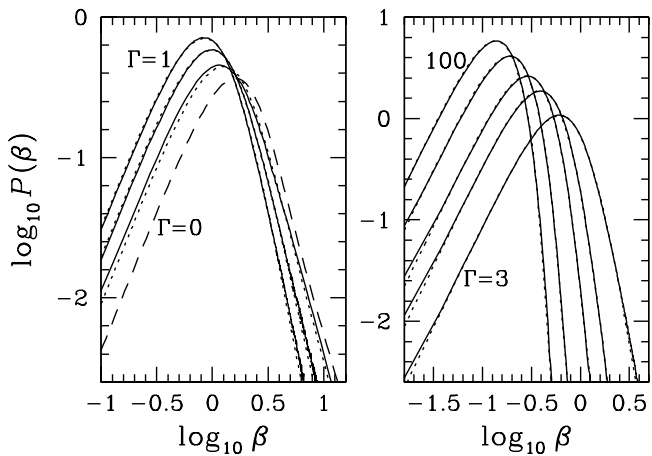}
     \end{center}
\caption{
The same as in Fig.~\protect\ref{fig-microf0}
but on the logarithmic scale.
}
\label{fig-microf0l}
\end{figure}

\subsubsection{Charged point}

For the charged test particle, the asymptotic behavior of $P(\beta)$
at large $\beta$ is qualitatively different in the cases
of zero and non-zero $\Gamma$.
When $\Gamma\neq0$, the
power-law decrease of $P(\beta)$ is replaced by
an exponential.
At moderate $\beta$ and large $\Gamma$,
$P(\beta)$ is approximately described by
 the Mayer distribution (\ref{Mayer}).
The cumulative function of this distribution is
\begin{equation}
   Q_M(\beta,\Gamma)=\textrm{erf}\left(\beta\,\sqrt{\Gamma/2}\right) -
      \sqrt{2\Gamma\over\pi}\,\beta\,e^{-\Gamma\beta^2 /2},
\label{Q_M}
\end{equation}
which is easily calculated using, e.g., the highly accurate
rational approximation to $e^{x^2} \textrm{erfc}(x)$ in Ref.~\cite{Press}.
The most probable value of the microfield provided by this distribution
is $\beta_m^M = \sqrt{2/\Gamma}$.
With increasing $\Gamma$, $\beta_m$ for a charged point decreases
faster than $\beta_m^\textrm{neu}$ given by Eq.~(\ref{beta_m-neu}).
The MC results for $\beta_m$ are approximately described
by the following modification of $\beta_m^M$:
\begin{equation}
  \beta_m^\textrm{ch} = \sqrt{2/\Gamma_\textrm{eff}},
\qquad
   \Gamma_\textrm{eff}\approx0.774+\Gamma^{1/4}+\Gamma.
\label{beta_m-ch}
\end{equation}

The microfield probability density at various values of $\Gamma$
is shown in Fig.~\ref{fig-microf0}.
The exponential decrease of $P(\beta)$
at large $\Gamma$ is more easily seen in the logarithmic
scale (Fig.~\ref{fig-microf0l}).
According to Eq.~(\ref{Pasymp}), the exponent at large $\beta$
is proportional to $\sqrt{\beta}$, and not to $\beta^2$
as in Eqs.~(\ref{Mayer}) and (\ref{Q_M}).
Nevertheless,
it is still possible to construct a self-consistent
analytic approximation
to $Q(\beta)$ and $P(\beta)$ analogous to Eq.~(\ref{fit-micron}).
Since the asymptotes are qualitatively different
in different coupling regimes, this approximation is more complicated:
\begin{equation}
  Q(\beta)
   =
    {Q_0(\beta)
      + 0.873\,\sqrt{\Gamma}\,Q_M(\beta,\Gamma_\textrm{eff})
   \over
    1+0.873\,\sqrt{\Gamma}},
\label{fit-microf}
\end{equation}
where
\begin{widetext}
\begin{equation}
   Q_0(\beta)
    =
   {
   q\,\beta^3\,\exp(-\Gamma' \beta^{1/2})+\beta^6
\over
    \left[\vphantom{\beta^{3/2}}
      2.25\pi\,q\,\left(1+\Gamma^{0.6}\right)^{-2.75} + 15.3\,\beta^2
   +1.238\,q\,\beta^3+\beta^{9/2}
      \right]\,\exp(-\Gamma' \beta^{1/2})+\beta^6
}\,,
\end{equation}
\end{widetext}
\[
    q
    =
       9.19 + 2.178\,\Gamma^{1.64},
\quad \mbox{and} \quad
   \Gamma'= \Gamma/(1+0.19\,\Gamma^{0.627}).
\]
At $\Gamma=0$, $\partial Q_0/\partial\beta$ reproduces $P_H(\beta)$
within 1\%.
The accuracy deteriorates at $\Gamma\approx0.1$ but improves again
at higher $\Gamma$: in Figs.~\ref{fig-microf0} and \ref{fig-microf0l},
differences between the fit and MC results
are barely visible at $\Gamma=0.5$, 3, 50, and 100.

\subsection{Effective screened potential}
\label{sect-Debye}

In this section, we consider microfield distribution
in an electron-screened Coulomb plasma.
Assuming that the ions interact via the effective potential
(\ref{V}), we have performed MC simulations
of $P(\beta)$ for various values of the screening parameter
from $s=0.05$ to $s=3.0$ and the Coulomb coupling parameter
from $\Gamma=0.1$ to $\Gamma=100$.
In accord with an intuitive expectation,
the MC simulations show that the
typical fields applied to a test particle are reduced
when the electron screening is taken into account.
In this case, the probability density $P$
and its cumulative function $Q$ depend on three
dimensionless arguments: $\beta$, $\Gamma$, and $s$.
Naturally, analytic approximations in this three-dimensional
space become complex and less accurate than
at fixed $s=0$; nevertheless we have attempted to
construct unified formulas for evaluation of $P(\beta)$
with an accuracy which is sufficient for most applications;
the results are presented below.

\subsubsection{Neutral point}

The most probable field strength applied to a neutral test particle,
evaluated by the MC method, can be parameterized as
\begin{equation}
   \beta_m^\textrm{neu}(\Gamma,s)\approx
      \beta_m^{(0)}(s)\,{1+0.15\,\sqrt{\Gamma} \over
         1+0.77\,(1+s)\,e^{-1.5s}\,\sqrt{\Gamma}},
\label{beta_neu_scr}
\end{equation}
where
\begin{equation}
   \beta_m^{(0)}(s) \approx [0.622 + 0.25\,s\,e^s]^{-1}.
\label{beta0}
\end{equation}
Equation (\ref{beta_neu_scr}) extends Eq.~(\ref{beta_m-neu})
to the case where $s\neq0$.
It is valid for the whole considered range of plasma parameters,
$\Gamma\leq100$ and $s\leq3$.

\begin{figure}
    \begin{center}
    \leavevmode
    \epsfxsize=86mm
    \epsfbox[145 280 460 590]{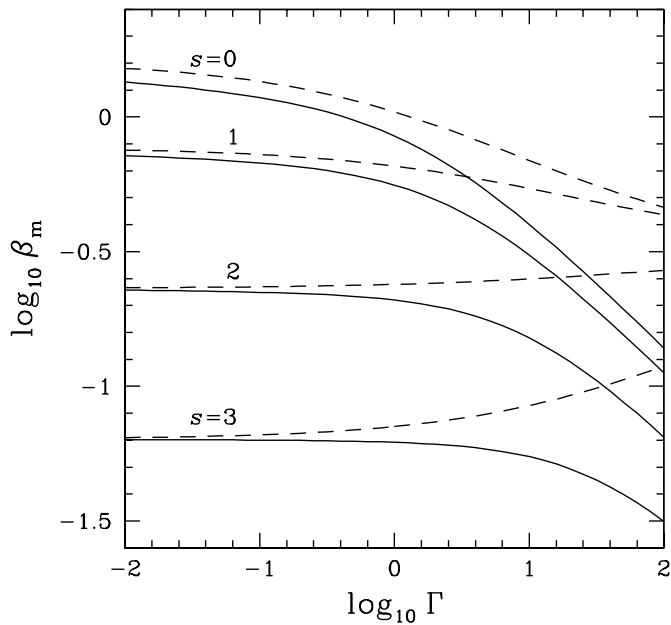}
    \end{center}
\caption{
The ``most probable'' field strength [normalized
according to Eq.~(\protect\ref{param})] as a function
of the Coulomb coupling parameter $\Gamma$
at four values of the screening parameter $s$.
Dashed lines: neutral point;
solid lines: charged point.
}
\label{fig-beta_m}
\end{figure}

\begin{figure}
    \begin{center}
    \leavevmode
    \epsfxsize=86mm
    \epsfbox[145 280 460 590]{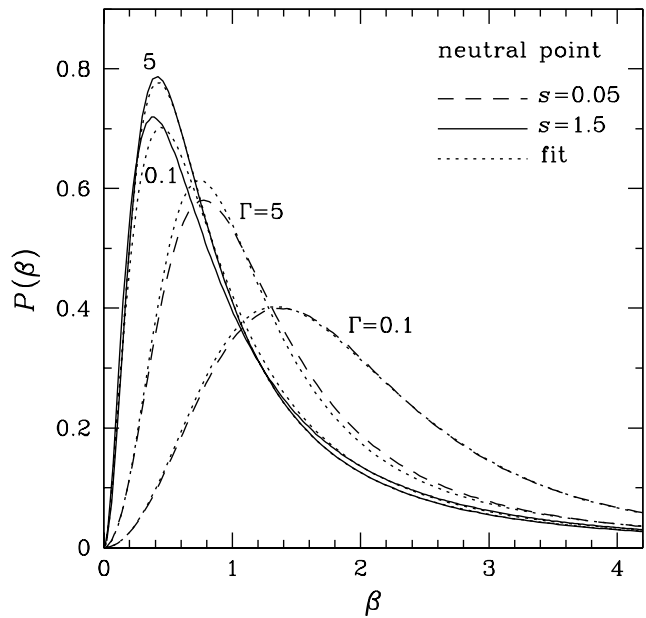}
    \end{center}
\caption{
Microfield distributions produced
at a neutral point in a plasma by ions
interacting via the screened Coulomb potential
for two values of the screening parameter $s$
(dashed lines: $s=0.05$; solid lines: $s=1.5$)
and two values of the Coulomb coupling parameter $\Gamma$.
The analytic approximation (Appendix \protect\ref{sect-A})
is shown by dotted lines.
}
\label{fig-micron1}
\end{figure}
The reduction of the typical microfield strength
applied to a charged test particle,
when the screening is taken into account, is illustrated
in Fig.~\ref{fig-beta_m}, where the dashed lines
correspond to the case of a neutral test particle,
which we consider in this section.
At a constant $\Gamma$, $\beta_m$ is a monotonically decreasing
function of $s$.
The dependence of $\beta_m$ on $\Gamma$ is less obvious.
At small values of $s$, $\beta_m$ decreases
monotonically with increasing $\Gamma$.
However, the opposite is observed when $s$ is large:
in this case, $\beta_m$ increases with growing $\Gamma$.
This implies that the most probable field strength
depends on $s$ stronger at small $\Gamma$ and weaker
at large $\Gamma$.

\begin{figure}
    \begin{center}
    \leavevmode
    \epsfxsize=86mm
    \epsfbox[145 280 460 590]{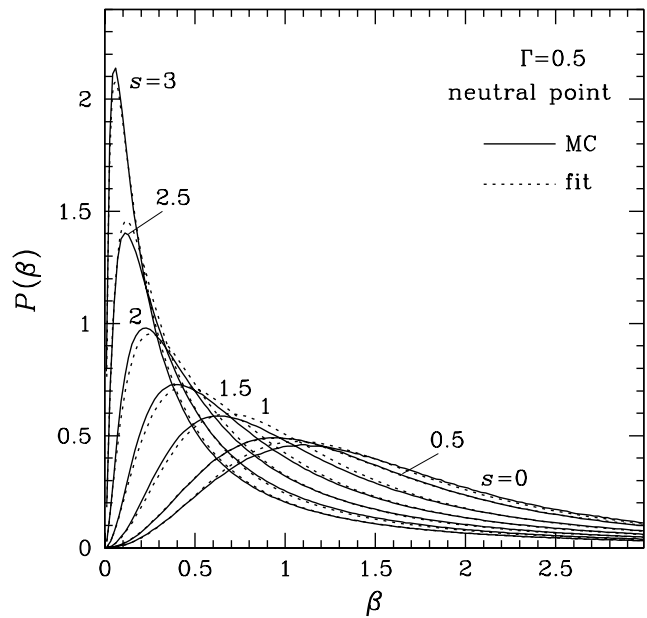}
    \end{center}
\caption{
Microfield distributions produced
at a neutral point for the coupling parameter $\Gamma=0.5$,
for 7 values of the screening parameter $s$
marked near the curves. MC results (solid lines)
are compared with the analytic approximation (dotted lines).
}
\label{fig-micron3}
\end{figure}

\begin{figure}
    \begin{center}
    \leavevmode
    \epsfxsize=86mm
    \epsfbox[145 280 460 590]{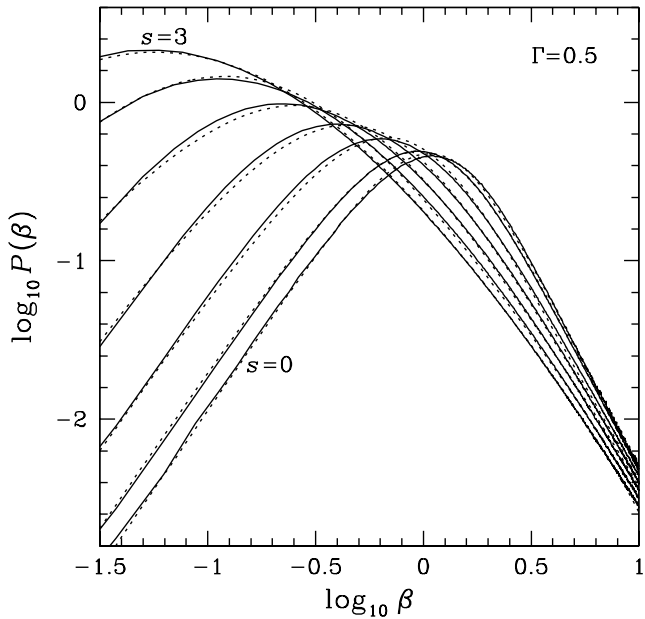}
    \end{center}
\caption{
The same as in Fig.~\protect\ref{fig-micron3}
 on the logarithmic scale.
}
\label{fig-micron3l}
\end{figure}

\begin{figure}
    \begin{center}
    \leavevmode
    \epsfxsize=86mm
    \epsfbox[145 280 460 590]{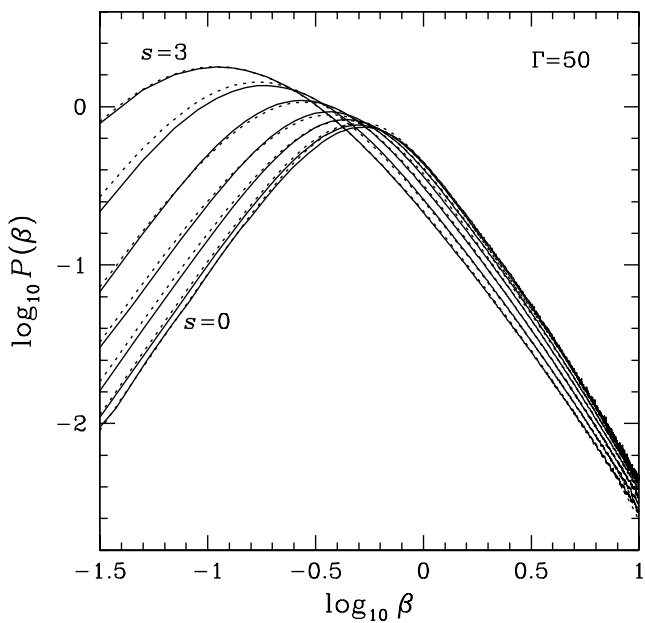}
    \end{center}
\caption{
The same as in Fig.~\protect\ref{fig-micron3l}
but for $\Gamma=50$.
}
\label{fig-micron2l}
\end{figure}

The modification of the probability density profile
with variation of the plasma parameters $\Gamma$ and $s$
is shown in Figs.~\ref{fig-micron1}--\ref{fig-micron2l}.
In the most important range of the coupling parameter,
$\Gamma\lesssim10$,
the dependence of $P(\beta)$ on $\Gamma$ becomes slow,
as the screening becomes sufficiently strong.
For example, in Fig.~\ref{fig-micron1} we observe a significant
modification of $P(\beta)$ at $s=0.05$, when
$\Gamma$ increases from $0.1$ to 5,
whereas $P(\beta)$ is only slightly modified
with the same increase of $\Gamma$, if $s=1.5$.

On the other hand, the profile of $P(\beta)$ strongly depends on
the value of the screening parameter $s$
(Fig.~\ref{fig-micron3}), especially if $s\gtrsim1$.
In this case, the most probable field is reduced drastically;
simultaneously, the distribution acquires a long ``tail,''
which shows that the values of $\beta\gg\beta_m$
occur more often than in the $s=0$ case.
Figure~\ref{fig-micron3l}, which presents the same dependences
on the logarithmic scale, clearly reveals
the two limiting power laws, $P\propto\beta^2$ and $\beta^{-5/2}$
at small and large $\beta$ values, respectively,
in agreement with Sec.~\ref{sect-asymp}.
We see that
these limits are approached considerably more slowly
 in the case of strong screening.

A comparison of Figs~\ref{fig-micron3l} and \ref{fig-micron2l}
shows that $P$ is less sensitive to $s$,
when $\Gamma$ is large, in accord with
the aforementioned property of $\beta_m$.
Nevertheless, even at $\Gamma\gg1$,
the sensitivity of $P$ with respect to $s$ remains essential.

It is possible to construct a fitting formula to
the probability function $Q(\beta)$ analogous to Eq.~(\ref{fit-micron}),
taking into account the screening.
Such parametric approximation is given in Appendix \ref{sect-A}
and is shown
in Figs.~\ref{fig-micron1}--\ref{fig-micron2l}
by dotted lines.
Although less accurate than
the fits presented in Sect.~\ref{sect-Coulomb},
this approximation reproduces well
the numerical results obtained by the MC simulation.

\subsubsection{Charged point}
\label{sect-charged_s}

Consider now the microfield distribution
created by plasma ions at a point where one of these ions
is placed, assuming that the Coulomb interaction is screened
according to Eq.~(\ref{V}).
As well as in the case of a neutral point,
the screening lowers typical microfield values.
Dependence of the most probable field $\beta_m$
on $s$ and $\Gamma$ can be approximated by
a simple expression
\begin{equation}
   \beta_m^\textrm{ch}(\Gamma,s) = \beta_m^{(0)}(s)
      \left[  1 + {\Gamma^{1/4} + \Gamma
         \over 0.774+0.54\,s\,e^s } \right]^{-1/2},
\label{beta_m_ch_s}
\end{equation}
where $\beta_m^{(0)}(s)$ is given by Eq.~(\ref{beta0}).
At $s=0$, Eq.~(\ref{beta_m_ch_s}) reproduces Eq.~(\ref{beta_m-ch}).
The dependence of $\beta_m$ on $\Gamma$
at various values of $s$ is plotted in Fig.~\ref{fig-beta_m}
by solid lines. Unlike the case of a neutral point
considered in the preceding section,
$\beta_m^\textrm{ch}$ decreases monotonically
with increasing $\Gamma$ at any given value of $s$.

\begin{figure}
    \begin{center}
    \leavevmode
    \epsfxsize=86mm
    \epsfbox[145 280 460 590]{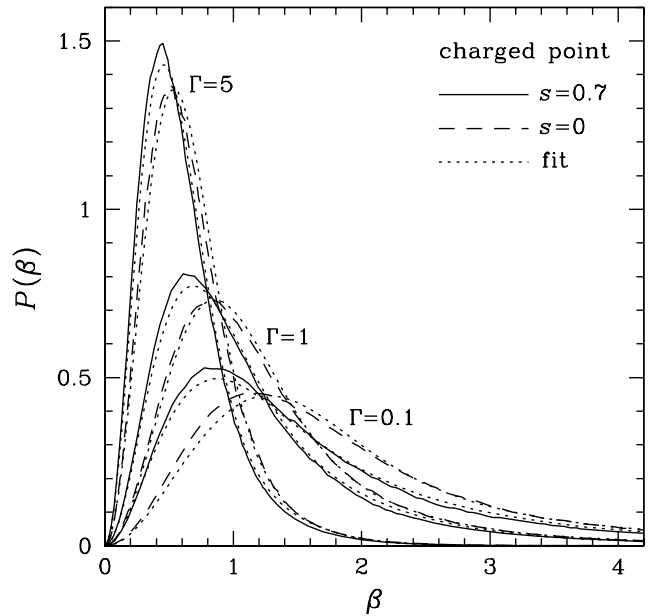}
    \end{center}
\caption{
Microfield distributions produced
at a charged test particle by ions
interacting via the unscreened (dashed lines)
and screened
Coulomb potential (solid lines; the screening parameter $s=0.7$),
for $\Gamma=0.1$, 1, and 5.
The analytic approximation given in Appendix~\protect\ref{sect-B}
is also shown for all cases (dotted curves).
}
\label{fig-microu1}
\end{figure}

\begin{figure}
    \begin{center}
    \leavevmode
    \epsfxsize=86mm
    \epsfbox[145 280 460 590]{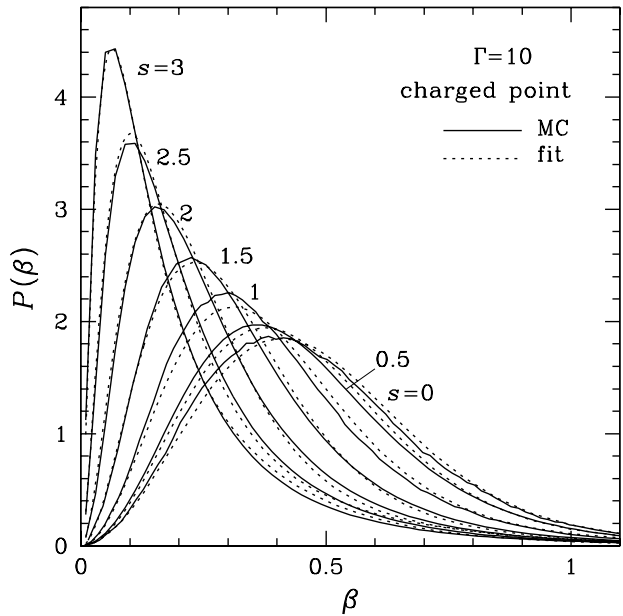}
    \end{center}
\caption{
Microfield distributions produced
at a charged test particle for the coupling parameter $\Gamma=10$,
for the values of the screening parameter $s$,
marked near the curves. MC results (solid lines)
are compared with the
analytic approximation given  in Appendix~\protect\ref{sect-B}
 (dotted lines).
}
\label{fig-microu2}
\end{figure}

\begin{figure}
    \begin{center}
    \leavevmode
    \epsfxsize=86mm
    \epsfbox[145 280 460 590]{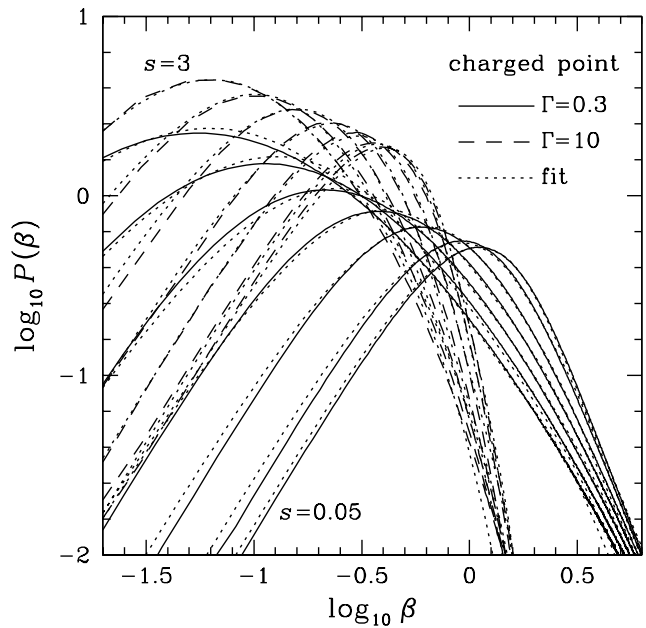}
    \end{center}
\caption{
Comparison of distributions $P(\beta)$
at two values of $\Gamma$ (solid lines: $\Gamma=0.3$;
dashed lines: $\Gamma=10$) and seven values of $s$
($s=0.05$, 0.50, 1.04, 1.5, 2.0, 2.5, and 3.0,
from right to left) on the logarithmic scale.
Dotted lines show the approximation (Appendix~\protect\ref{sect-B}).
}
\label{fig-micro23l}
\end{figure}

\begin{figure}
    \begin{center}
    \leavevmode
    \epsfxsize=86mm
    \epsfbox[145 280 460 590]{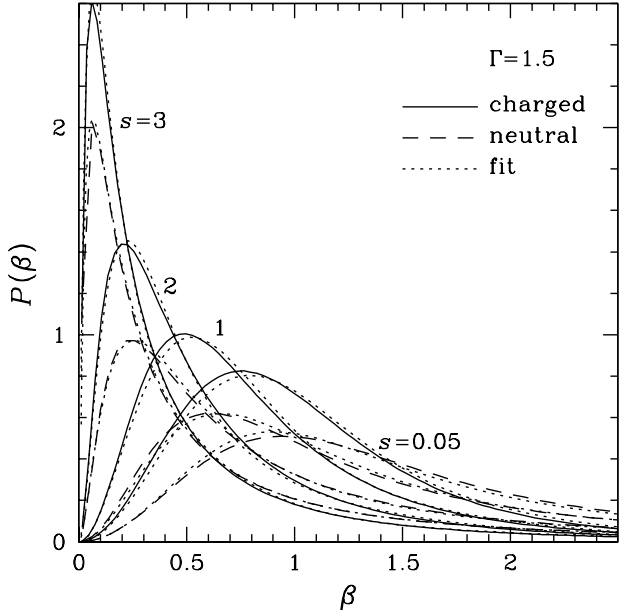}
    \end{center}
\caption{
Comparison of the probability density distributions $P(\beta)$
at a neutral (dashed lines) and charged (solid lines) point
at $\Gamma=1.5$ and for four values of $s$.
Dotted lines show the approximations in Appendixes~\protect\ref{sect-A}
and \protect\ref{sect-B}.
}
\label{fig-micronu1}
\end{figure}

The modification of the probability density
$P(\beta)$ with increasing $s$ or $\Gamma$
is illustrated by Figs.~\ref{fig-microu1}--\ref{fig-micronu1}.
The dependence on the coupling parameter $\Gamma$
remains qualitatively the same as without screening:
with the increase of $\Gamma$,
the typical field strengths become lower,
and the distribution $P(\beta)$ becomes narrower.
The increase of $s$ also shifts the peak of $P$ to smaller $\beta$.
At $s>1$, however,
the latter shift is accompanied by a striking modification of
the shape of the function $P$:
a fast growth at small $\beta$ is followed by
a slow, gradual decrease at $\beta>\beta_m$.

As in the case of the Coulomb potential,
the limiting behavior of $P(\beta)$ at $\beta\gg\beta_m$
changes from power law at $\Gamma=0$
to the exponential decrease
$\propto\beta^{-5/2}e^{-\Gamma\sqrt{\beta}}$
at $\beta\to\infty$. This limiting law
is reached very slowly, if $s$ is large,
as clearly seen in the logarithmic scale
(Fig.~\ref{fig-micro23l}).
When $\beta$ is moderately large
and $\Gamma\gg1$, the decrease is approximately Gaussian,
$\propto e^{-\Gamma\beta^2/2}$.

The rich variety of the shapes of $P(\beta)$
(depending on $s$ and $\Gamma$)
complicates significantly the construction
of a fitting formula. In this case,
a unified fit to $Q$ and $P$ (like those
presented in the previous sections)
would become too cumbersome.
Therefore we have chosen to construct an analytic
approximation to the function $P$ only. Whenever necessary,
$Q$ can be found by numerical integration [Eq.~(\ref{Q})].


Our approximate formula is presented in Appendix \ref{sect-B}.
Its quality is revealed by Figs.~\ref{fig-microu1}--\ref{fig-micronu1},
where the fit is compared with results of the MC simulations.
The typical accuracy of several percent at $s\lesssim1.5$
is expected to be sufficient for most applications.
At $s > 1.5$ the accuracy deteriorates, and at $s>2$
the asymptotic behavior at large $\beta$ is not reproduced,
as one can see from the logarithmic plots (Fig.~\ref{fig-micro23l}).
At such strong screening, the fit still may be used for
evaluation of $P(\beta)$ not too far from $\beta_m$.
Indeed, on the linear scale (Figs.~\ref{fig-microu2},
\ref{fig-micronu1}) the difference between the fit
and the MC results appears to be small even at $s=3$.

Figure \ref{fig-micronu1} allows one to compare
the screening effects in the two cases of a neutral and
charged test particle.
Since $\Gamma$ is greater than unity,
the difference between $P(\beta)$ functions in the
two cases is large at $s=0.05$
(i.e.,  for a nearly Coulomb potential),
in agreement with Sec.~\ref{sect-Coulomb}.
With the increase of $s$, however, the difference
becomes smaller near the peak of $P(\beta)$.
The positions of the peaks for a neutral and charged points
almost coincide
at $s > 1$, and the difference in their heights
is caused by the fact that, in the case of a neutral point,
 $P(\beta)$ decreases much more slowly at large $\beta$,
and therefore the region of $\beta\gg\beta_m$
gives a larger contribution to the normalization integral,
than it does in the case of a charged point.

\section{Summary}
\label{sect-end}
We have calculated
microfield distributions at neutral and charged test particles
in a one-component plasma of ions, interacting
via Coulomb potential, in various regimes
from weak to strong coupling.
The MC and APEX methods of calculation yield similar
distributions,
in agreement with previously known results
\cite{APEX}.
Self-consistent elementary-function approximations
for the field probability density $P(\beta)$
and its cumulative distribution $Q(\beta)$ are constructed
in the two cases of a neutral and charged point,
for a Coulomb coupling parameter $\Gamma$ varying
from 0 to $10^2$.

Furthermore, MC calculations of the microfield distribution
have been performed for the screened Coulomb interaction,
using the model of ions interacting via the Debye-like (Yukawa)
effective potential, with an effective screening length
as a second independent parameter.
The dimensionless screening parameter $s$
[Eq.~(\ref{param})] varies from 0 to 3.
The whole set of numerical results for $P(\beta)$
at various values of the coupling and screening parameters
is approximated by analytic expressions.

The obtained results
can be used in theoretical models of optical spectra
and equations of state of Coulomb plasmas.

\begin{acknowledgments}
A.~P.\ thanks V.\ S.\ Filinov for
useful discussion and the theoretical astrophysics group
at the Ecole Normale Sup\'erieure de Lyon for hospitality.
The work of A.~P.\ has been partially supported by
RFBR Grant No.\ 99-02-18099.
\end{acknowledgments}

\appendix
 \vspace*{5ex}

\section{High-field asymptotic expression
for the microfield distribution}
\label{sect-C}

In this Appendix, we outline the derivation of the strong-field
asymptotic limit of Eq.~(\ref{P-gen}).
For brevity, we adopt the convention that all lengths and radius vectors
are measured in units of the ion-sphere radius $a$.

We take advantage of the fact that,
for Coulomb interactions, high-field contributions are produced
by nearest particles. This well-known result has been
investigated in Ref.~\cite{Hooperasymp}.
We assume (i) that
the microfield at $\bm{r}_0=\bm{0}$ is dominated by the contribution
of the nearest neighbor ion located at $\bm{r}_1$, and
(ii) that only the potential of this ion contributes to the potential energy.
Then Eq.~(\ref{P-gen}) can be approximated by
\begin{widetext}
\begin{equation}
  P(\beta) \sim P_\textrm{as}(\beta) = 4\pi \beta^2\;
  { \int_\Omega d\bm{r}_1\int_{r_2 > r_1}d\bm{r}_2
    \ldots \int_{r_N > r_1} d\bm{r}_N\;
          e^{-\Gamma v(r_1)} \delta(\bm{\beta} - \nabla v(r_1))
\over
\int_\Omega d\bm{r}_1\int_{r_2 > r_1}d\bm{r}_2 \ldots
    \int_{r_N > r_1} d\bm{r}_N\; e^{-\Gamma v(r_1)}
}.
\end{equation}
\end{widetext}
Here, $\Omega$ is the total volume of the system,
and $v(r)=V(r)\, a/(Ze)^2$ is the reduced potential,
so that $\Gamma v(r)= V(r)/k_B T$.
In our units, $\Omega=\frac{4}{3}\pi N$.

Taking into account that $\int_\Omega d\bm{r}=\Omega$,
$\int_{r_2 > r_1}d\bm{r}_2=\Omega (1-{r_1^3 / N})$,
and $\lim_{N\to\infty} (1-r_1^3/N)^N = e^{-r_1^3}$,
we obtain
\begin{equation}	
  P_\textrm{as}(\beta)  = \beta^2\;
  { \int_\Omega
          e^{-r^3-\Gamma v(r)} \delta(\bm{\beta} - \nabla v(r))
          \,d\bm{r}
\over
   \int_0^\infty r^2 e^{-\Gamma v(r)}\,dr }\,.
\end{equation}

The coordinate transformation
$\bm{u}= \nabla v(|\bm{r}|) = \tilde\beta({r})\, \bm{r}/r$
with the Jacobian $J= (r^2/\tilde\beta^2) |dr/d\tilde\beta|$ yields
\begin{equation}
  P_\textrm{as}(\beta)= {1\over K}\, r^2
  \left|{dr \over d\tilde\beta}\right|_{_{\tilde\beta=\beta}}\,
  e^{-r^3-\Gamma v(r)}
\end{equation}
with $\beta=\tilde\beta(r)=|{dv(r) / dr}|$ and $K=\int_0^\infty
        r^2 e^{-r^3-\Gamma v(r)}\,dr$.

For a Coulomb potential, $v(r)={1/r}$ and $\beta ={1/r^2}$,
so that
\begin{equation}
   P_\textrm{as}(\beta) \sim {\beta^{-5/2}\,
   \exp(-\Gamma\beta^{1/2}-\beta^{-3/2})
\over
          2\int_0^\infty{\beta'}^{-5/2}\,
             \exp(-\Gamma{\beta'}^{1/2}-{\beta'}^{-3/2})\,d\beta'} .
\label{Paris}
\end{equation}

For a Yukawa potential $v(r)=e^{-s r}/r$, we have
\begin{subequations}
\label{Yukas}
\begin{equation}
  P(\beta)= {1\over K} {r^2 \,\exp[-r^3-\Gamma {e^{-s r}/ r}]
\over
(2+{s r}+ {2/s r}) \,(s/  r^2)\, e^{-s r}},
\end{equation}
where $r$ should be determined from the equation
\begin{equation}
\beta= (1 + s r) \frac{e^{-s r}}{r^2},
\end{equation}
and
\begin{equation}
   K=\int_0^\infty r^2 \exp[-r^3-\Gamma {e^{-s r}/ r}]\,dr.
\end{equation}
\end{subequations}
Equation (\ref{Yukas}) is considerably more complicated than Eq.~(\ref{Paris});
it can be compared to Hooper's formulation
\cite{Hooperasymp}. On the other hand, the simpler Eq.~(\ref{Paris})
or its generalization, Eq.~(\ref{Pasymp}), can be sufficiently accurate
for most applications. This has been verified
by comparing
with MC results \cite{Gilles97,Gh95} and is illustrated
in Fig.~\ref{fig-microMCapexASYMP}.

\section{Approximation to the probability
function of microfields at a neutral point}
\label{sect-A}

In this Appendix we present a fitting formula
to the probability function $Q(\beta)$, Eq.~(\ref{Q}),
for a neutral point in a plasma with Coulomb coupling
and screening.

At every pair of $\Gamma$ and $s$ values, we
derive a Pad\'e approximation to the
microfield probability function,
\begin{equation}
   Q(\beta)= {a_0 \beta^3 - 2 \, \beta^{9/2} + \beta^6
     \over
 a_1 + a_2 \beta + a_3 \beta^2 + a_4 \beta^3 - \beta^{9/2} + \beta^6}\,.
\label{fit-micron_s}
\end{equation}
This expression ensures that, when its derivative is taken,
the limits $P(\beta)\propto \beta^2$ at $\beta\to 0$
and  $P(\beta)\approx 1.5\beta^{-5/2}$ at $\beta\to \infty$
are reproduced.
At arbitrary $\beta$, an agreement with MC results
is provided by an appropriate choice
of the fitting parameters $a_0$--$a_4$.
The latter parameters, in turn, can be approximated
as functions of $\Gamma$ and $s$:
\begin{eqnarray}
   a_0 &=& {97\,s^2+1.29\,s^7 \over 1+3.1\times10^{-3}\,s^5}
      +(59 + 8.1\,s^2)\,g,
 \\&&
      g\equiv\sqrt{0.08+\Gamma},
 \nonumber\\
   a_1 &=& {1.16\over 1+0.188\,s^6}
      \left[ 1+ {103\,g^{\alpha(s)} \over 1+0.33\,s} \right],
 \\&&
     \alpha(s) \equiv {0.068+0.038\,s^7 \over 1+0.030\,s^7},
 \nonumber\\
   a_2 &=& {95\,s \over 1+6\times10^{-3}\,s^7} + 1.2\,s^2\,g,
 \\
   a_3 &=& 27\,s^3 + 36\,g,
 \\
   a_4 &=& {1.894 + s \over 2 + s}\, a_0.
\end{eqnarray}

This approximation has been checked for the whole range of
the plasma parameters for which the MC simulations were performed,
i.e., at $0\leq\Gamma\leq100$ and $0\leq s\leq3$.
In the case of purely Coulomb potential ($s=0$), however,
Eq.~(\ref{fit-micron}) should be used as more accurate.

\section{Approximation to the probability
density of microfields at a plasma ion}
\label{sect-B}

In this Appendix we present an analytic approximation
to the probability density $P(\beta)$ of electric microfield
at an ion in an electron-screened Coulomb plasma.

At every $\Gamma$ and $s$,
we write
\begin{equation}
   P(\beta) \approx {\beta^2\over S_N}
   \left[ A\,e^{-a\beta^\alpha} + B\,e^{-b \beta^\gamma} +
    {e^{-\Gamma\,\beta^{1/2}} \over 1+c\,\beta^{9/2}}
\,\right],
\label{fit-microu}
\end{equation}
where  $S_N$ is the normalization constant.
For the latter constant, we have
\begin{equation}
   S_N = A\,{\Gamma(3/\alpha)\over\alpha a^{3/\alpha}}
      + B\,{\Gamma(3/\gamma)\over\gamma b^{3/\gamma}}
      + \Gamma^{-6} F(c/\Gamma^9),
  \label{S_N}
\end{equation}
where $\Gamma(3/\alpha)$ and $\Gamma(3/\gamma)$
are the Gamma-function values which are easily calculated
(e.g., Ref.~\cite{Press}), and
\begin{equation}
  F(y)\equiv\int_0^\infty {x^2\,e^{-\sqrt{x}}
     \over 1+ y x^{9/2}}\,dx.
\label{S2}
\end{equation}
For the latter integral, we have constructed
an approximation,
\begin{eqnarray}&& \hspace*{-3em}
   F(y) \approx \left(1+{4\pi\over9\sqrt{3}}\,y^{1/9} \right)
\nonumber\\&&\hspace*{-3em}  \times
    \left[ \frac{1}{240} + 0.849\, y^{1/3}
   \! + 3.2 \, y^{5/9}
    \! +2.43 \, y^{2/3}
    \! + y^{7/9} \right]^{-1}\hspace*{-1em} .
\label{S2fit}
\end{eqnarray}
Expression (\ref{S2fit}) fits the integral (\ref{S2})
within 0.8\%. This is sufficient for evaluation of
$Q(\beta)$ and $P(\beta)$ in most applications.
The accuracy of Eq.~(\ref{S2fit}) may be insufficient, however,
if the values of $[1-Q(\beta)]$ at $\beta\gg\beta_m$
are of interest. In this case,
the normalization constant can be evaluated numerically
according to Eqs.~(\ref{S_N}) and (\ref{S2}).

Equation (\ref{fit-microu}) ensures that $P(\beta)\propto\beta^2$
at $\beta\to0$. Moreover, it also ensures the correct limiting
behavior at $\beta\to\infty$, Eq.~(\ref{Pasymp}), with $\tilde\Gamma=\Gamma$,
provided that $\alpha$ and $\gamma$ are both greater than 0.5.
With the choice of parameters presented below, this is the case
at $0\leq s \lesssim 2.3$ (any $\Gamma$),
which covers the whole range of values
 typically encountered in stellar
and laboratory dense plasmas.

We have parameterized $A$, $a$, $\alpha$, $B$, $b$, and $\gamma$
in Eq.~(\ref{fit-microu}) as functions of $\Gamma$,
having the same form at any $s$:
\begin{eqnarray}
   A &=& A_1\,{1+A_4\,\sqrt{\Gamma}
       \over 1 + A_2\,\Gamma^2 + A_3\,\Gamma^4},
   \\
   a &=& a_0 + \Gamma/2,
   \\
   \alpha &=& {\alpha_1 + 2\,\alpha_2\,\sqrt{\Gamma}
          \over 1 + \alpha_2\,\sqrt{\Gamma}},
 \\
   B &=& {B_1
       \over 1 + B_2\,\Gamma^2 + B_3\,\Gamma^4},
   \\
   b &=& b_0 + \Gamma/4,
   \\
   \gamma &=& {\gamma_1 + 1.5\,\gamma_2\,\sqrt{\Gamma}
          \over 1 + \gamma_2\,\sqrt{\Gamma}}.
\end{eqnarray}
 Then the parameters of these expressions
 ($A_n$, $B_n$, $a_0$, $b_0$, $\alpha_1$,
 $\alpha_2$, $\gamma_1$, and $\gamma_2$)
 and the parameter $c$ of Eq.~(\ref{fit-microu})
  have been approximated by analytic functions of $s$:
\begin{subequations}
\begin{eqnarray}
   A_1 &=& 0.59 + 2540\,s^4 + 3\,s^{14},
   \\
   A_2 &=& {0.55+10\,s^{0.5} + 2\,s^{4.5}
      \over 1+20\,s^{0.5}},
   \\
   A_3 &=& 2.17\times10^{-3}\,s^5,
   \\
   A_4 &=& 14.8/[1+117\,s^{3.5}],
\end{eqnarray}
\end{subequations}
\begin{eqnarray}
   a_0 & =& 1.15+2\,s^{1.8},
\end{eqnarray}
\begin{subequations}
\begin{eqnarray}
   \alpha_1 &=& 0.1 + 1.1/(1+0.145\,s^3),
   \\
   \alpha_2 &=& {5.4 \over 1+20 s^2} + {1.1 \over 1+14 s^{0.35}},
\end{eqnarray}
\end{subequations}
\begin{subequations}
\begin{eqnarray}
   B_1 &=& 0.386 + 300\,s^2 + 1.1\,s^{9.5},
   \\
   B_2 &=& 0.038+0.79\,s^{0.75},
   \\
   B_3 &=& {3.7\times10^{-3}\,s^{5.5}
      \over 1+4\times10^{-3}\,s^9},
\end{eqnarray}
\end{subequations}
\begin{eqnarray}
   b_0 &=& (1+0.54\,s^{2.5})/(1+0.07 s),
\end{eqnarray}
\begin{subequations}
\begin{eqnarray}
   \gamma_1 &=& 0.1 + 1.1/(1+0.174\,s^{2.5}),
   \\
   \gamma_2 &=& {5.4 \over 1+21 s^{1.5}} + {1.1 \over 1+19 s^{0.16}},
\end{eqnarray}
\end{subequations}
\begin{eqnarray}
   c &=& {0.097\over 1+210\,s^{2.5} \,\exp(-1.3\,s^{1.5})}.
\end{eqnarray}
The high powers of $s$ in some of these equations
effectively describe the strong $s$-dependence
of the shape of the function $P(\beta)$ at $s\gtrsim1$.

As noted in Sec.~\ref{sect-charged_s},
this approximation is valid at $\Gamma\leq100$ and $s\leq2$,
but it can be also used at $2<s\leq3$, provided that $\beta$
is not larger than $\approx10\beta_m^\textrm{ch}$.

\end{document}